\documentclass[preprint2]{aastex}
\usepackage{graphicx}
\usepackage{natbib}

\def\d{\,d$^{-1}$}
\def\mh{\,$\mu$Hz}

\def\ds{$\delta$\,Scuti}
\def\te{$T_{\mathrm{eff}}$}

\def\sun{\hbox{$\odot$}}

\slugcomment{accepted by ApJL}

\begin{document}

\title{Evidence for granulation in early A-type stars.}

\author{Thomas Kallinger\altaffilmark{1, 2} and Jaymie M. Matthews\altaffilmark{1}}

\altaffiltext{1}{Department of Physics and Astronomy, University of British Columbia, 6224 Agricultural Road, Vancouver, BC V6T 1Z1, Canada}
\altaffiltext{2}{Institute for Astronomy, University of Vienna, T\"urkenschanzstrasse 17, 1180 Vienna, Austria}

\begin{abstract}
Stars with spectral types earlier than about F0 on (or close) to the main sequence have long been believed to lack observable surface convection, although evolutionary models of A-type stars do predict very thin surface convective zones. We present evidence for granulation in two \ds\ stars of spectral type A2: HD\,174936 and HD\,50844. Recent analyses of space-based CoRoT (Convection, Rotation, and planetary Transits) data revealed up to some 1000 frequencies in the photometry of these stars. The frequencies were interpreted as individual pulsation modes. If true, there must be large numbers of nonradial modes of very high degree $l$ which should suffer cancellation effects in disk-integrated photometry (even of high space-based precision). The p-mode interpretation of all the frequencies in HD\,174936 and HD\,50844 depends on the assumption of white (frequency independent) noise. Our independent analyses of the data provide an alternative explanation: most of the peaks in the Fourier spectra are the signature of non-white granulation background noise, and less than about 100 of the frequencies are actual stellar p-modes in each star. We find granulation time scales which are consistent with scaling relations that describe cooler stars with known surface convection. If the granulation interpretation is correct, the hundreds of low-amplitude Fourier peaks reported in recent studies are falsely interpreted as independent pulsation modes and a significantly lower number of frequencies are associated with pulsation, consistent with only modes of low degree. 
\end{abstract}

\maketitle

\section{Do more frequencies actually mean more modes?}

The \ds\ stars are intermediate-mass pulsating variables with spectral types A $-$ F in the lower part of the classical instability strip. They are opacity-driven pulsators, which can exhibit large numbers of excited acoustic modes, making them key candidates for asteroseismology. Because of this modeling potential, huge efforts have been made to observe \ds\ stars from ground, offering increasingly rich eigenfrequency spectra, thanks to steady improvements of observational techniques. The pinnacle of this attempts was the star FG Vir, for which \cite{bre05} detected 75 frequencies with amplitudes down to $\sim$200\,ppm. The trend of finding oscillations of lower and lower amplitude led to the belief that, only the tip of the eigenspectrum iceberg was visible and as noise levels were pushed lower by better observations, more modes would be found in most pulsating stars.

This has been impressively confirmed by recent space-based observations. \cite{mat07} reported on MOST \citep[Microvariability \& Oscillation of STars;][]{wal03, mat04} observations of HD\,209775, revealing a similar number of frequencies as had been found in FG Vir. But truly groundbreaking results came from CoRoT \citep{bag06}, which uncovered an unexpectedly rich eigenspectrum in the \ds\ star HD\,50844, with potentially more than 1000 frequencies \citep{por09}. Similar was found for HD\,174936 \citep{gar09}, and results on other \ds\ stars in the CoRoT sample are currently processed. The basic result is that several hundreds of frequencies have been detected in several of the \ds\ stars monitored with CoRoT. 

This early result inspired a new perspective on \ds\ pulsation, potentially addressing long-standing problems of the mode amplitude distribution and unknown mode selection process(es) in \ds\ stars. But at the same time, it also opened new fundamental problems: how can such large numbers of modes be actually detected in measurements of integrated light from stars? Asteroseismic models predict some tens of acoustic modes (maybe one hundred if rotationally split frequencies are included in the tally) with degrees $l \leq 4$, which are potentially excited in the frequency range of {\ds}-type oscillations. A substantially larger number of modes than this would have to include modes of degree $\ell > 4$ whose net amplitudes should suffer strong cancellation effects when integrated across the stellar disk \citep[see, e.g., ][]{daz06}. We find it implausible that hundreds of modes with high degree $l$ would survive amplitude cancellation, even with space-based precision, unless some effective spatial filter was present on the stellar disk. High-contrast large-scale structures on the stellar surface could act as such a spacial filter. Such features are known for active cool stars (and for strongly magnetic B $-$ F stars) but have not yet been observed among \ds\ stars. We prefer a simpler interpretation for the multitude of frequencies in the newly observed power spectra of \ds\ stars: non-white stellar background noise, due to surface granulation.

Stars in the \ds\ part of the HRD are believed to have no measurable surface convection. Because of this long-standing view, space-based photometry of \ds\ stars has been analysed assuming the noise is dominated by white (frequency-independent) photon noise. The CoRoT light curves of HD\,174936 and HD\,50844 were processed using standard iterative sine-wave fitting algorithms. One of them, SigSpec \citep{reg07}, assumes the variance in the time series to be dominated by white noise and rates the significance of the signal components according to this assumption. This approach clearly fails in the case of frequency-dependent noise. The noise peaks in the frequency range of an enhanced noise can have much larger amplitudes than the real (white) noise and are potentially misinterpreted as significant signal. Other tools, like Period04 \citep{len05}, are less restrictive and can handle frequency-dependent noise. On the other hand, the program has technical difficulties to simultaneously fit several hundred signal components and is therefore used to check/confirm the main frequencies (up to say 100 in number) identified by other programs.

Non-white intrinsic stellar noise is a feature of cool stars with convective envelopes. The turbulent motions in the stars' outer convective envelopes generate quasi-stochastic power acting on various time scales, with amplitudes strongly decreasing with increasing frequency. Although the signal is stochastic, it exhibits distinctive characteristics. The Fourier transform of the signal follows simple power laws where different physical processes on (or near) the stellar surface produce the same type of signal but on different time and amplitude scales. For the Sun, the different signal components are usually assigned to stellar activity, activity of the photospheric/chromospheric magnetic network, and granulation, with time scales ranging from months (for active regions) to minutes (for granulation) \citep[see e.g.,][]{mic09}. This has also been seen in other Sun-like stars \citep[see e.g.,][]{mic08}, and also for the more evolved red giants \citep[see e.g.,][]{kal09}.

The presence of a granulation background signal depends of course on the presence of a surface convection zone $-$ a property which is usually attributed to cool stars beyond the red border of the instability strip. But according to stellar evolutionary models, stars located within the instability strip have a thin, but nevertheless non-negligible, convective surface layer (see Sect.\,3). The proper modeling of convection is still a weak facet of most stellar evolution calculations. But the existing models give reason to consider that a subtle granulation background signal should be observable in A stars. We believe that this was not observed (at least to our knowledge) and recognised before because the amplitudes are very low, requiring precise, long, and uninterrupted observations to detect the signal. Granulation amplitudes are measured in power density because the actual amplitude in the observations depend on the sampling and length of the time series. For the Sun, granulation has an amplitude on the order of less than 1\,ppm$^2$/{\mh} \citep{mic09} and typically 10$^3$ to 10$^4$\,ppm$^2$/{\mh} for stars on the lower red-giant branch \citep{kal09}. We expect the amplitudes for \ds\ stars to be somewhere in between.


\section{Power spectrum modeling: Stellar background and pulsations}

For the Sun, it is common practice to model the background signal with a sum of power laws, $P(\nu) = \sum_i a_i  / (1 + (\nu /  b_i)^{c_i})$, where $\nu$ is the frequency, $a_i$, $b_i$, and $c_i$ are the amplitude, characteristic frequency (inverse time scale), and slope of the power laws, respectively. The number of components usually ranges between two and five, depending on the frequency coverage of the observations. Power-law models were first introduced by \citet{har85} with the slope fixed to 2. But \citet{aig04} and \citet{mic09} have shown that, at least for the Sun, the true slope is closer to 4. The presence of an additional signal, like solar-type p modes, will significantly distort the fit, and one has to include a term in a model to account for that additional signal. \citet{kal09} have approximated the shape of the pulsation power excess in red giants by a Gaussian, which is also reasonable for main-sequence solar-type pulsators \citep{gru09}. This is more complicated for \ds\ stars as the excess envelope can be quite different from a Gaussian. We therefore pre-whiten the largest amplitude frequencies and assume the residual power excess to be at least Gaussian-like. For a detailed frequency analysis a more realistic model might be needed. But this is not our current goal.								

\begin{figure*}[t]
\centering
\includegraphics[width=1.0\textwidth]{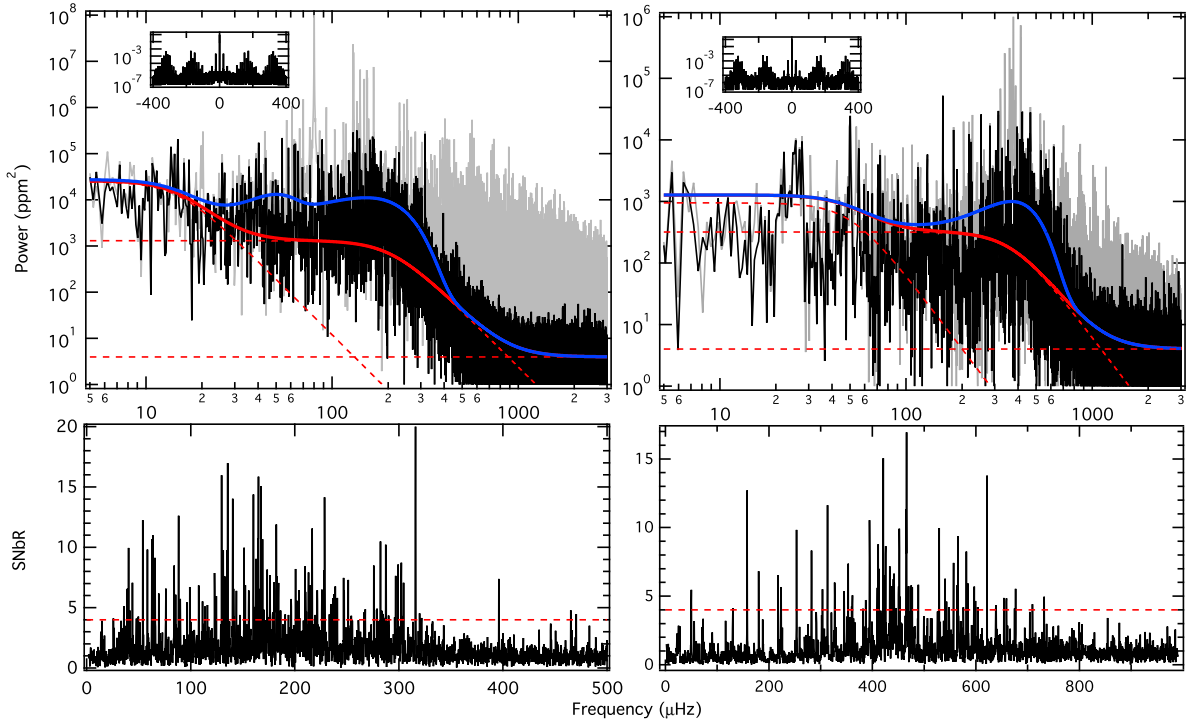}
\caption{\textit{top panels:} Original (grey curve) and residual gap-filled (black curve) spectra of HD\,50844 (right) and HD\,174936 (left) after pre-whitening the 20 and 10 most significant frequencies, respectively, and a global model fit (blue curve -- in the online version only). Red lines represent the models without the pulsational components consisting of two power laws (dashed lines) and white noise. \textit{bottom panels}: Residual power spectra normalised to the power-law components. Dashed lines indicate a signal-to-background-noise ratio of SNbR = 4.0.}
\label{fig:fourier}
\end{figure*}

Following \citet{kal09}, we model the residual power spectra with a superposition of white noise, a sum of power laws, and a Gaussian.

\begin{equation}
P(\nu) =  P_n +  \sum_{i=1}^2 \frac{a_{i}}{1 + (\nu / b_i)^4} + P_g \cdot e^{\,\,-(\nu_g - \nu)^2 / (2\sigma_g^2)},
\label{eq:fit}
\end{equation}

\noindent where $P_n$ represents the white noise contribution and $P_g$, $\nu_g$, and $\sigma_g$ are the height, central frequency, and width of the power excess hump, respectively. We use a Bayesian Markov-Chain Monte Carlo algorithm to fit the global model to the residual power spectra. Details about the fitting procedure may be found in \citet{gru09}.

\subsection{HD\,174936}

The field \ds\ star HD\,174936 (A2; V = 8.6) was observed within the CoRoT asteroseismology program during the first 27-day short run campaign. A first analysis was presented by \citet{gar09}, who detected about 420 individual frequencies with SNR $\ge 4.0$, assuming white noise. This corresponds to a detection limit of about 14\,ppm.

We use SigSpec to pre-whiten the ten most significant frequencies from the CoRoT time series, which are identical to those reported by \citet{gar09}. Although the star was monitored continuously, a significant number of data points are flagged during the reduction process \citep{sam07}, mainly outliers due to high-energy particle impacts during the satellite's passes through the South Atlantic Anomaly. Like \citet{gar09}, we simply remove these questionable data points, degrading the duty cycle of the time series to $\sim$90\%. The resulting gaps are fairly regular, producing low-amplitude aliasing peaks in the power spectrum. This is not a problem for a pre-whitening sequence but distorts a direct fit to the spectrum. We therefore fill the gaps by linear interpolation, which is the standard procedure for solar-type pulsators \citep[see e.g.][]{apo08}. The power spectra of the original data and of the pre-whitened and gap-filled data are plotted in Fig.\,\ref{fig:fourier}, with the spectral window of the original data shown in the inset. The gap-filling suppresses any alias signal and obviously changes the power spectrum dramatically for frequencies above $\sim$600\mh . Note that such gap-filling might influence the determinations of individual frequencies and amplitudes, which is not our goal in this study. 

We then fit a global model (Eq.\,\ref{eq:fit}) to the residual power spectrum. In the case of HD\,174936, the residual {\ds}-type oscillation power appears to be concentrated in a single power excess centered on about 350\mh . The resulting global fit reproduces the overall shape of the residual power spectrum and demonstrates the strong frequency dependence of the background signal. At low frequencies, we find the background signal to be as high as about 36\,ppm. It decreases to $\sim$19\,ppm at 100\mh\ before it fades into the white noise of $\sim$2\,ppm. 

Our next step is to use the power-law components of the fit to correct the power spectrum for the frequency-dependent background noise. The resulting signal-to-background-noise ratio (SNbR) is given in Fig.\,\ref{fig:fourier}. We find 55 frequencies exceeding a significance limit of SNbR $\ge 4.0$ resulting in a total of 65 significant frequencies, which we attribute to {\ds}-type pulsations. This is small compared to the 420 pulsation modes identified by \citet{gar09}, but in much better agreement with the maximum number of frequencies reasonably expected for low-degree p-modes in a \ds\ star.

HD\,174936 is a metal-deficient star with an iron abundance $[Fe/H] \simeq -0.32$ \citep{cha06}. We have determined an effective temperature $ T_{\rm eff}$ = 8080$\pm$160\,K from Str\"omgren photometry \citep{hau98} using the \citet{mon85} calibration, which gives also an absolute visual magnitude $M_V$ = 1.91$\pm$0.32. With $M_{bol,\sun}$ = 4.75 and a bolometric correction $BC_V \simeq 0.003$ \citep{lej01}, we obtain $L/L\sun = 14.4 \pm 4.4$ and $R/R\sun$ = 1.91$\pm$0.35.

\subsection{HD\,50844}

HD\,50844 (A2; V = 9.09) was continuously observed by CoRoT during the 58-day initial run. A detailed frequency analysis and first asteroseismic interpretation was performed by \cite{por09}. In their analysis, they must pre-whiten the time series with more than 1000 individual frequency terms to decrease the residual rms in the time series to what they believe is the detection limit of the data. From their Fig.\,3, it is obvious that they would need yet thousands more frequencies to bring the mean amplitude at frequencies below 50 \d\ to the mean amplitude at high frequencies. This alone is a strong argument for the presence of strongly non-white noise. 

Following the approach described above, we pre-whiten the 20 most significant frequencies from the CoRoT data of HD\,50844 and fill the gaps in the residual time series by linear interpolation. Unlike HD\,174936, the oscillation power of HD\,50844 seem to be distributed over a broader frequency range, with two clusters around about 50 and 180\mh . We therefore include a second Gaussian in our global model of the power spectrum (Eq.\,\ref{eq:fit}). The power spectrum of the original data is shown in Fig.\,\ref{fig:fourier} along with the power spectrum of the pre-whitened and gap-filled data, as well as the most probable global model with and without the Gaussian terms. In the case of HD\,50844, the contrast between the low-frequency background signal and the high-frequency white noise is even higher than for HD\,174936. We find a low-frequency background level of about 160\,ppm, which decreases to $\sim$35\,ppm at 100\mh , before it fades into the high-frequency white noise of $\sim$2\,ppm. (Coincidentally, the white noise component for HD\,50844 is almost the same as for HD\,174936. This is because the average photon flux of the HD\,50844 observations is smaller than for HD\,174936, but the time series consists of more than twice as many data points.) 

We find 105 frequencies exceeding a SNbR $\ge$ 4.0, resulting in a total of 125 significant frequencies which we attribute to {\ds}-type pulsations, compared to more than 1000 suggested by \cite{por09}.

HD\,50844 is also a metal-poor star, slightly more evolved than HD\,174936, with $T_{\rm eff}$ = 7500$\pm$200\,K, $M_V$ = 1.31, and $[Fe/H] \simeq -0.4$ \citep{por05}. With $BC_V \simeq 0.043$ \citep{lej01}, we obtain $L/L\sun$ = 24$\pm$8 and $R/R\sun$ = 2.9$\pm$0.5.

\begin{figure}[t]
\centering
\includegraphics[width=0.5\textwidth]{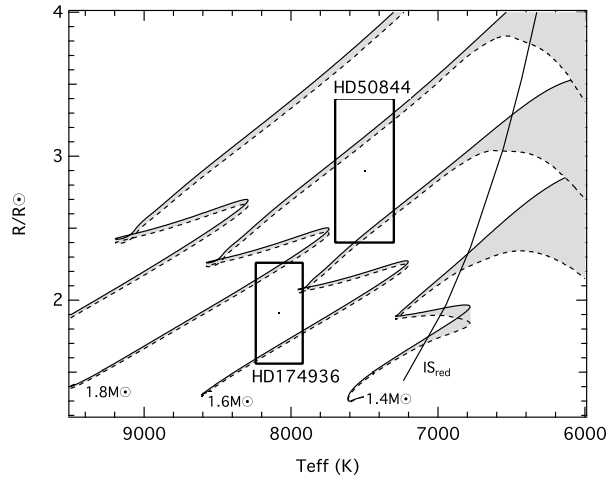}
\caption{R versus \te\ for metal-poor YREC models. The radii of the bases of the surface convection zones (grey-shaded area) are shown as dashed lines.}
\label{fig:model}
\end{figure}

\section{Depth of the convection zone and granulation time scales}

In Fig.\,\ref{fig:model}, we compare the effective temperatures and radii of HD\,174936 and HD\,50844 with those of metal-poor evolutionary tracks, which were computed with the Yale Stellar Evolution Code \citep[YREC;][]{gue92,dem08} for a initial chemical composition (Y, Z) = (0.28, 0.01) and a mixing length parameter $\alpha$ = 1.8. Details about the model physics adopted can be found in \citet{kal08} and references therein. We estimate the masses of HD\,174936 and HD\,50844 from their positions in this diagram to be 1.7$\pm$0.2 and 1.75$\pm$0.25\,M\sun, respectively, which agrees with the findings of \citet{gar09} and \citet{por09} using a similar method. 

More important in this context, Fig.\,\ref{fig:model} also shows the radius of the base of the surface convection zone. Our intermediate-mass models have a very thin outer convection zone ($\sim$1\,\% in radius) during their main sequence evolution. But it starts to slowly expands after core hydrogen depletion in the H-shell burning phase before the base of the convection zone rapidly drops near the red border of the instability strip and the models become almost fully convective on the giant branch (outside the plotted range). The exact position of the base of the convection zone will almost certainly depend on the actual treatment of convection in the models and, in our case, on $\alpha$. But the general picture is clear. Stars located in the \ds\ regime of the HRD are expected to have thin, but non-negligible, outer convective envelopes and therefore should exhibit a surface granulation signal. We find that the outer convection zone could be as deep as $\sim$1 and 1.5\% of R for HD\,174936 and HD\,50844, respectively.

One might intuitively expect that the granulation amplitude should be very low for such thin convective layers. But in fact the photometric amplitude of the granulation signal is not directly correlated with the depth of the convection zone. It can be assumed that the granulation amplitude is inversely proportional to the total number of granulation cells on the stellar surface, where the average cell size is believed to be proportional to the atmospheric pressure scale height $H_p$ \citep[see e.g.][]{ste07}. The granulation time scales, on the other hand, are believed to scale with the ratio of the cell size and cell velocity, where the cells are assumed to move at speeds proportional to the local sound speed $c_s$. Under the assumption of an ideal adiabatic gas, the granulation frequency (inverse time scale), $\nu_{gran}$, should scale as $\nu_{gran} \propto c_s / H_p$ \citep[see e.g.][and references therein]{hub09}. Unlike \ds\ stars, solar-type pulsators show very regular patterns of oscillation modes almost symmetrically distributed around the so-called ``frequency of maximum oscillation power'', $\nu_{max}$. According to \citet{kje95}, $\nu_{max}$ is also proportional to $c_s / H_p$, and is believed to scale as $M/(R^2 \sqrt{T_\mathrm{eff}})$. This suggests that the inverse granulation time scales determined from our global model fits scale as

\begin{equation}
\nu_{gran} \propto \nu_{max} \propto M \cdot R^{-2} \cdot T_\mathrm{eff}^{-1/2}
\label{eq:fit}
\end{equation}

\noindent with the fundamental parameters in solar units. 

\begin{figure}[t]
\centering
\includegraphics[width=0.5\textwidth]{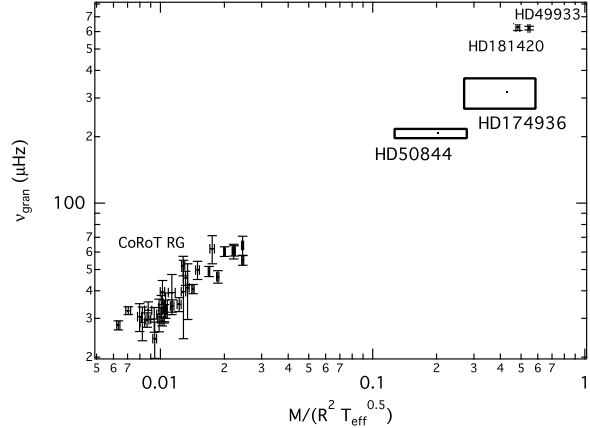}
\caption{The characteristic $\nu_{gran}$ as a function of the $\nu_{max}$ for main-sequence and red-giant solar-type pulsator. Parameters on the abscissa are given in solar units and show how $\nu_{max}$ scales with the stellar parameters.}
\label{fig:timescale}
\end{figure}

This relation is shown in Fig.\,\ref{fig:timescale}, where we compare the granulation frequencies with $\nu_{max}$ of a sample of main-sequence and red-giant solar-type pulsators \citep{kal09}. All parameters have been determined directly from CoRoT data using our global model approach. For the two \ds\ stars we cannot of course measure $\nu_{max}$ as they do not pulsate in solar-type p-modes of high radial order. But we can estimate from their fundamental parameters what would be their $\nu_{max}$ if solar-type oscillations were excited in them, and then compare these virtual frequencies to the measured granulation frequencies. And indeed, the two stars follow the expected linear relation. We therefore conclude that the granulation time scales are consistent with those expected from a simple scaling relation for other stars unambiguously showing surface granulation, and that we have indeed measured a granulation signal. 

\section{Conclusions and future prospects}

We have shown that the multitude of low-amplitude frequencies found in the CoRoT observations of \ds\ stars are consistent with a strongly frequency-dependent intrinsic background signal. Such a signal is well known for main-sequence and red-giant stars, but has never been recognised in \ds\ stars. We have fit the power spectra of two such stars with a simple model, allowing us to measure the stars' granulation time scales and to disentangle the granulation signal from the pulsation signal. Consequently, we find a much smaller number of frequencies that we attributed to {\ds}-type pulsation than the hundreds to a thousand individual modes reported by others for the same stars. Our number of pulsation modes is more consistent with the number of frequencies that can be expected from integrated light observations of low-degree p modes. We have estimated granulation time scales for the two stars which are consistent with those expected from a simple scaling relation for stars known to exhibit surface granulation. 

For HD\,174936, we find a low-frequency background of $\sim$3000\,ppm$^2$/{\mh}, which corresponds to $\sim$36 ppm in the Fourier amplitude spectrum of the 27-day-long CoRoT run. This should be roughly the same for a MOST run of comparable duration, and there are several \ds\ stars which have been monitored by MOST for this long. We plan to reanalyse the MOST light curves of bright \ds\ stars to check for granulation signatures.  For one year of Kepler data, we expect a signal of $(27/365)^{1/2} \times$ 36\,ppm $\simeq$ 10\,ppm. For HD\,50844, the low-frequency signal is $\sim$125,000\,ppm$^2$/{\mh}, which corresponds to $\sim$160\,ppm in the 58-day CoRoT run. For a 30-day MOST run, we estimate a low-frequeny background of 220\,ppm, and for a one-year Kepler run, 60\,ppm.

Our result has shown that with the increasing quality of space-based photometry, intrinsic {\em frequency-dependent} noise can no longer be neglected in the frequency analyses of \ds\ stars. The traditional method to ascribe significances to candidate pulsation frequencies assuming white noise in a \ds\ star poses the danger of over-interpretating the data. This is especially important in the context of the upcoming Kepler observations which will also include observations of \ds\ stars, but with even higher precisions and longer time coverages than possible with MOST and CoRoT.

\end{document}